\documentclass[3p,times,twocolumn]{elsarticle}

\usepackage{ecrc}

\volume{00}

\firstpage{1}

\journalname{Nuclear Physics B Proceedings Supplement}

\runauth{}

\jid{nuphbp}

\jnltitlelogo{Nuclear Physics B Proceedings Supplement}

\usepackage{amssymb}
\usepackage{subfigure}

\usepackage[figuresright]{rotating}
\usepackage{pdflscape}
\usepackage{longtable}
\usepackage{subfigure}
\usepackage{multirow}
\usepackage{color}
\usepackage{colortbl}
\usepackage{ctable}
\usepackage{xspace}

\begin{document}

\begin{frontmatter}



\dochead{}

\title{The origin of Cosmic-Rays from SNRs: confirmations and challenges after the first direct proof.}


\author[Arc]{M.Cardillo\corref{cor2}}
\ead{martina@arcetri.astro.com}
\author[CNR,Uni2,CIFS]{M.Tavani}
\author[Mi]{A.Giuliani}

\cortext[cor2]{Principal corresponding author}

\address[Arc]{INAF-Osservatorio Astrofisico di Arcetri,
Largo E.Fermi 5, 50125, Florence (Italy)}
\address[CNR]{INAF/IAPS, I-00133 Roma, Italy}
\address[Uni2]{Dip. di Fisica, Univ. Tor Vergata, I-00133 Roma, Italy}
\address[CIFS]{CIFS-Torino, I-10133 Torino, Italy}
\address[Mi]{INAF/IASF-Milano, I-20133 Milano, Italy}

\begin{abstract}

Until now, providing an experimental unambiguous proof of Cosmic Ray (CR) origin has been elusive. The SuperNova Remnant (SNR) study showed an increasingly complex scenario with a continuous elaboration of theoretical models.
The middle-aged supernova remnant (SNR) W44 has recently attracted attention because of its relevance regarding the origin of Galactic cosmic-rays. The gamma-ray missions AGILE and Fermi have established, for the first time for a SNR, the spectral continuum below 200 MeV which can be attributed to neutral pion emission.
Our work is focused on a global re-assessment of all available data and models of particle acceleration in W44 and 
our analysis strengthens previous studies and observations of the W44 complex environment, providing new information for a more detailed modeling. 
However, having determined the hadronic nature of the gamma-ray emission on firm ground, a number of theoretical challenges remains to be addressed in the context of CR acceleration in SNRs.
\end{abstract}


\end{frontmatter}

\section{Introduction}
Cosmic-rays (CRs) are highly energetic particles (with kinetic energies up to $E=10^{20}$ eV) mainly composed by protons and nuclei with a small percentage of electrons (1$\%$). Since from their discovery Cosmic-Rays are one of the most debated issues of the
high energy astrophysics. Their origin is still a fundamental problem and is the subject of very intense research
 \cite{fermi49,ginzburg64,berezinskii90}, \cite[for recent reviews, see][]{helder12,aharonian13}. Focusing on CRs
produced in our Galaxy (energies up to the ``knee'', $E=10^{15}$~eV), strong shocks in Supernova Remnants (SNRs) are considered the most probable CR sources \cite[e.g.,][]{ginzburg64}, \cite[recent review in] []{vink12}.
However, the final proof for the origin of CRs up to the knee can only be obtained through two fundamental signatures.
The first one is the detection of a clear gamma-ray signature  of $\pi^{0}$ decay in Galactic sources; the second one is the identification of sources emitting a photon spectrum up to PeV energies. Both indications are quite difficult to obtain. The "Pevatron" sources are notoriously difficult to find \cite[for a review, see][]{aharonian13}, and the neutral pion decay signature is not easy to identify because of the possible contribution from co-spatial leptonic emission. Hadronic (expected to produce the $\pi^{0}$ decay spectral signature) and leptonic components can in principle be distinguished in the 50-200 MeV energy band, where they are expected to show different behaviors.\\
Over the last five years, AGILE and Fermi gamma-ray satellites, together with ground telescopes operating in the TeV energy range (HESS, VERITAS and MAGIC), collected a great amount of data from SNRs \cite{abdo09_W51,abdo10_CasA,abdo10_IC443,abdo10_W49b,abdo10_W44,abdo10_W28,abdo11_1713,
acciari09_IC443,tavani10_IC443,acciari10_CasA,acciari11_tycho,aharonian01_CasA,aharonian07_1713,aharonian08_W28,aleksic12_W51,giordano12_tycho,
giuliani10_W28,hewitt12_PuppisA,katsuta12_S147,lemoine12_RCW86} providing important information and challenging theoretical models. Most of the SNRs detected in the $\gamma$-ray band are interacting with a MC. The presence of a high density target, indeed, enhances the possibility to detect $\gamma$-ray emission from pp-interaction. Interestingly, most of the observed \rm SNRs show, apparently, a spectrum steeper than the one expected from linear and non-linear diffusive shock acceleration models (DSA) of index near 2, and possible convex spectrum \cite{bell78a,malkov01,blasi05}.\\
W44 is one of the most interesting SNRs observed so far. It is a middle-aged SNR, which is bright at gamma-ray energies and quite close to us.
Its gamma-ray spectral index is $p\sim3$ \cite{GiuCaTa11}, in apparent contradiction with DSA  models. Its environment is very interesting and complex, requiring a careful re-evaluation of theoretical models. Recently, an analysis of Fermi-LAT data confirmed these results \cite{ackermann13_W44}. 
Our new analysis of the AGILE data, comparing with the new Fermi-LAT data and considering also radio and new CO data from VLA and NANTEN2, confirms that gamma-ray data can be fitted only with a hadronic model and that the spectral index is $p\geq 3$. Moreover, even if there is an ambiguity on the choice of a specific hadronic model, we obtain some important constraints on fundamental parameters, such as the magnetic field and the ISM density \cite{cardillo14}.
Thanks to a deep analysis of available multiwavelength data, we collected multiwavelength information for young and middle-aged SNRs in order to compare their spectral behaviors with the W44 one and, in general, with the main theoretical model expectations.\\
In spite of the great amount of data from SNRs emission, the understanding of CR acceleration and propagation processes is all but complete.
\section{The supernova remnant W44}  
\label{SNRW44}

The SNR W44 is a middle-aged ($\sim$20,000 yrs old) SNR located in the Galactic Plane ($l,b$)=~($34.7,-0.4)$ at a
distance $d\sim3.1$~kpc \cite{clark76,wolszczan91}. 
Multiwavelength observations revealed interesting features. In the radio band, W44 shows a quasi-elliptical shell \cite[][and
references therein]{castelletti07}; the radio shell asymmetry is probably due to expansion in an inhomogeneous ISM. In the
northwest side of the remnant, which correlates with a peak of the radio emission, there is bright [SII] emission characteristic of shock-excited radiative filaments \cite{giacani97}. In the southeast side, instead, there is a molecular cloud (MC) complex embedded in the SNR shell that interacts with the source \cite{wootten77,rho94}. The OH maser (1720~MHz) emission detected in correspondence with the SNR/MC region, confirm their interaction \cite{claussen97,hoffman05}. In \cite{wolszczan91} the discovery of the radio pulsar PSR~B1853+01 is reported, which is
located in the south part of the remnant and surrounded by a cometary-shaped pulsar wind nebula (PWN) \cite{frail96}. This system, however, does not appear to be correlated with the detected gamma-ray emission. The X-ray observations of W44  by the Einstein Observatory \cite{watson83} showed centrally peaked emission, which is later confirmed by Chandra data \cite{shelton04}. \\
The SNR~W44 is well studied also in the gamma-ray band. In \cite{abdo10_W44}, a GeV morphology well correlated with the radio emission is showed, together with a steep photon spectrum (index near 3) that, however, has a low-energy threshold of 200 MeV, limiting the chance to identify a neutral pion signature. The relatively large  gamma-ray brightness of W44 and the good spectral capability of AGILE near 100 MeV
\cite{tavani09,vercellone08,vercellone09} have stimulated a thorough investigation of this supernova remnant with the AGILE data.
The AGILE gamma-ray spectrum in the range of 50 MeV to 10 GeV confirms the high-energy steep slope up to 10 GeV and, remarkably,
identifies a spectral decrease below 200 MeV for the first time, as expected from neutral pion decay \cite[][hereafter G11]{GiuCaTa11}.
In the analysis of G11, both leptonic and hadronic models were considered in fitting both AGILE and Fermi-LAT data. The best model was determined to be dominated by hadronic emission with a proton distribution of spectral index $p_{2}=3.0 \pm 0.1$ and a low-energy cut-off at $E_{c}= 6\pm 1$~GeV. The low-energy spectral behavior seen by AGILE was recently confirmed by the Fermi-LAT team that revisited the gamma-ray emission from W44 \cite[][hereafter A13]{ackermann13_W44}. Their best hadronic model with an \textit{assumed} surrounding medium density $n\sim100$ cm$^{-3}$ is based on a smoothed broken power-law hadronic distribution with a break energy $E_{br}=22$~GeV and indices $p_{1}=2.36$ for $E<E_{br}$ and $p_{2}=3.5$ for $E > E_{br}$. Model parameters in A13 differ from those considered earlier in \cite{abdo10_W44}. Apparently, bremsstrahlung
emission is not considered to be relevant in the hadronic modeling of A13, even though this process could provide a non-negligible contribution to the gamma-ray emissivity in principle.\\
We present here a new analysis of AGILE data with a revised assessment of the W44 surrounding environment, which is based on new CO data obtained from the NANTEN2 telescope \cite{cardillo14}.\\
The very important feature of the SNR~W44 spectrum, confirmed in every analysis, is its slope at GeV energies: the index $p\sim3$
is substantially steeper than the range that is plausibly expected in linear and non-linear DSA  models.
In \cite{malkov11_W44}, this spectral feature is explained by Alfv\'en damping in the presence of a relatively large-density medium where acceleration occurs. The W44 environment is quite challenging in its morphology and requires a reanalysis of its properties in the context of the crucial implications for the acceleration mechanism of CRs.

\section{New AGILE data analysis}
\label{newAGILE}

\begin{figure}[!ht]
\centering
\includegraphics[bb=0 -32 683 552,scale=0.32]{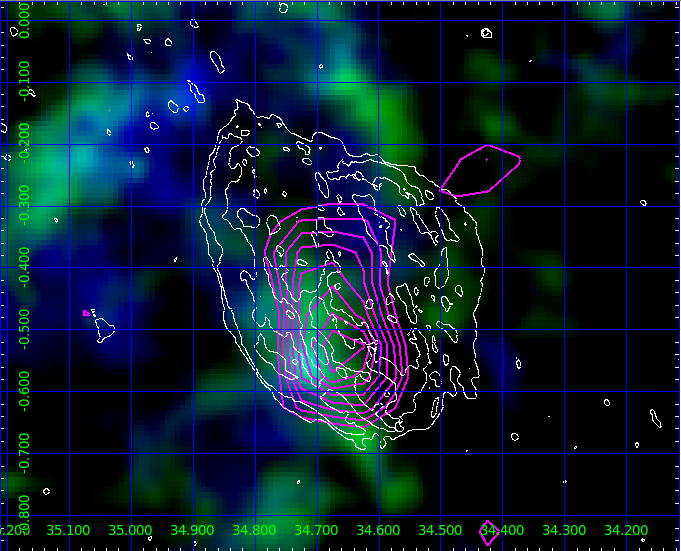}
\caption[width=.4\textwidth]{Combined CO data from the NANTEN2 observatory that is superimposed with the
AGILE gamma-ray data contours above 400 MeV of
the W44 region (map in Galactic coordinates) and VLA contours \cite{cardillo14}}
\label{maps}
\end{figure}

We performed a global reassessment of the AGILE  data on W44, including new gamma-ray data
obtained until June 2012 \cite{cardillo14}.
Fig.~\ref{maps} shows the W44 AGILE CO maps in two velocity channels, 41 and 43 km/s, with radio (VLA) and gamma-ray (AGILE, 400-10000 MeV). Gamma-ray emission appears to be mostly concentrated near a high-density region, the CO peak at (34.7,-0.5), indicating that most of the W44 gamma-ray emission is coincident with  a site of SNR/MC interaction \cite[For a more detailed description and image, see][]{cardillo14}.

\begin{figure}[!ht]
 \begin{center}
 \includegraphics[bb=0 0 192 135,scale=1.2]{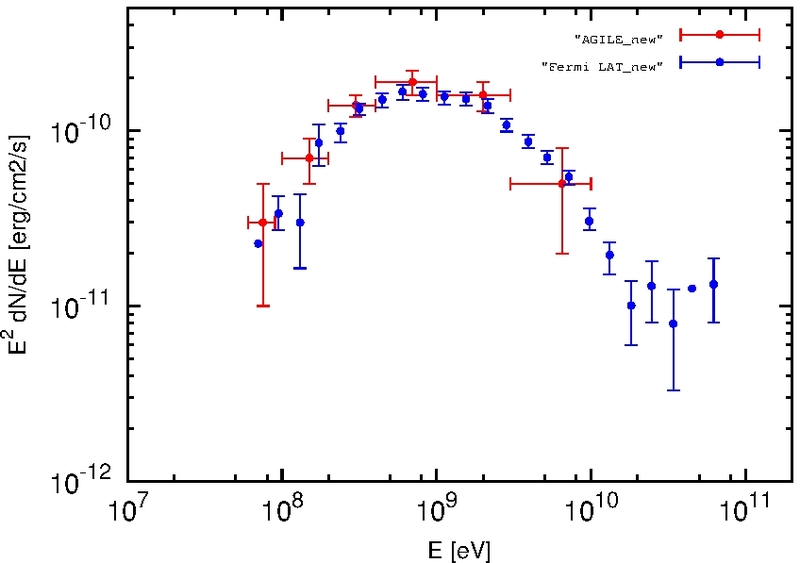}
\end{center}
\caption{AGILE new gamma-ray spectrum of SNR~W44 \cite{cardillo14}
superimposed with the Fermi-LAT data from \cite{ackermann13_W44}.}
\label{spectra}
\end{figure}

Figure~\ref{spectra} shows the AGILE gamma-ray spectrum with the
recently updated Fermi-LAT data from \cite{ackermann13_W44}.
The AGILE spectrum is composed by six energy bins between 50~MeV
and 10~GeV and our error-bars takes statistical errors into account. The measured flux of the source above 400 MeV is
$F=(23\pm2)\times10^{-8}$~ph~cm$^{-2}$~s$^{-1}$.
We notice the good agreement between the two spectra.
Especially important is the confirmation of the drastic spectral decrement below 200 MeV, a crucial feature that is discussed
below. Both AGILE and Fermi-LAT spectra differ from the previously published spectra in G11 and \cite{abdo10_W44}, \cite{cardillo14}.

\subsection{Modeling}
\label{modeling}
 
\begin{figure*}[!ht]
\centering
 \subfigure{\includegraphics[bb=0 0 873 609,scale=0.21]{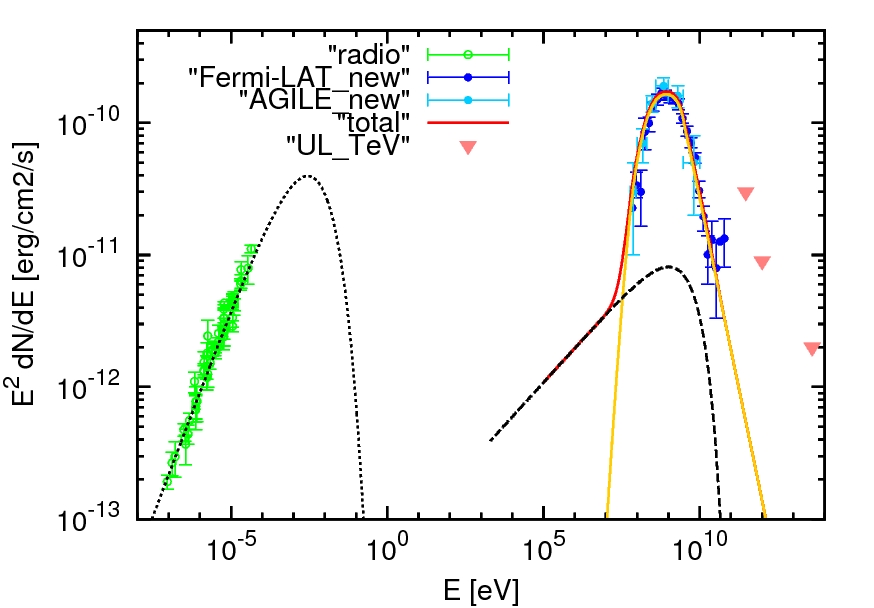}}\qquad
 \subfigure{\includegraphics[bb=0 0 871 607,scale=0.21]{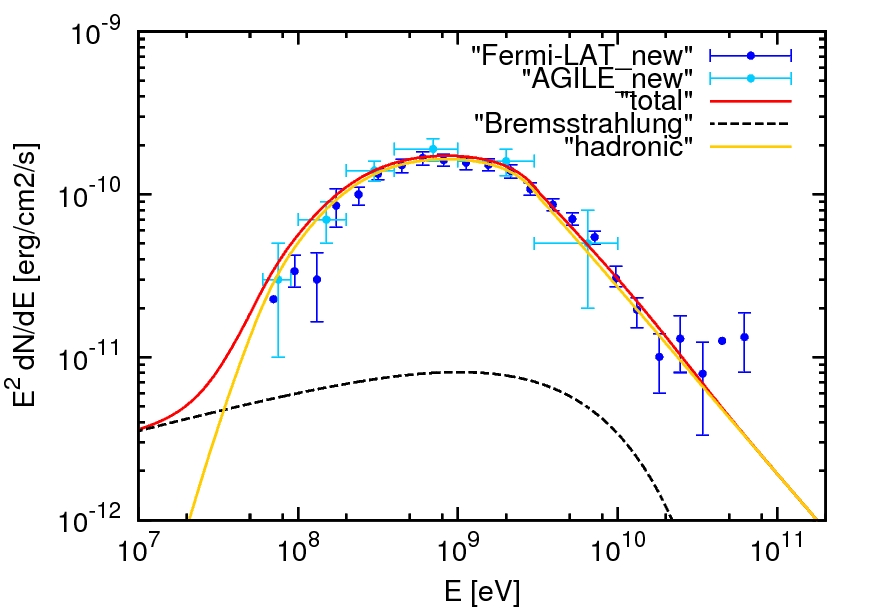}}
 \caption{Our best hadronic model, H3, of the broadband spectrum of the SNR~W44 that is superimposed with radio
 and gamma-ray data of Fig.~\ref{spectra}.
Proton distribution in Eq.~\ref{broken_2} with index
$p_{1}=2.2\pm0.1$ (for $E<E_{br}$) and $p_{2}=3.2\pm0.1$ (for
$E>E_{br}$) where \textit{$E^{p}_{br}$}= 20 GeV. This model is
characterized by \textit{B} =210~$\mu$G and \textit{n}
=300~cm$^{-3}$. We show the neutral pion emission
from the accelerated proton distribution discussed in the text.
The black curves show the electron contribution by synchrotron
(dot) and bremsstrahlung (dashed) emissions; the IC contribution is
negligible. The total gamma-ray emission from
pion-decay and bremsstrahlung is also shown. \textbf{(Left Panel)}: SED from
radio to gamma-ray band. \textbf{(Right Panel)}: only gamma-ray
part of the spectrum.}
\label{hadronic}
\end{figure*}
We model the radio, AGILE, and Fermi-LAT spectral data  by hadronic and leptonic-only scenarios by considering the new NANTEN2 CO data that provides a value for the ISM density in the SNR surroundings, $n_{av}\simeq 250\,\rm cm^{-3}$ \cite{yoshiike13_W44}. This value of
the average gaseous density that surrounds the gamma-ray emission is substantially larger than the one assumed in G11 and A13
($n=100 \, \rm cm^{-3}$). Since the AGILE gamma-ray emission is strongly correlated with one of the CO peaks, we consider an average density
$n\simeq 300\pm 50$ cm$^{-3}$ $>n_{av}$ in the following. In modeling the spectra, we consider the most statistically significant Fermi-LAT data up to 50 GeV.
We assume that the gamma-ray emission spectrum is due to the combined contribution of hadronic $\pi^{0}$ emission and leptonic bremsstrahlung emission by considering the proton component as the main one. For hadronic emission, we use the formalism explained in \cite{kelner06} that is a good approximation of the exact solution. We consider a proton distribution in total energy $E$ rather than in kinetic energy $E_{k}=E-m_{p}c^{2}$, following \cite{simpson83} and \cite{dermer86}, but with $\delta$-function approximation for the cross section \cite{aharonian04}..
We fit the gamma-ray data by assuming different types of proton distributions in energy: a simple power-law with a high-energy cut-off (\textbf{model H1}), a smoothed broken power-law (\textbf{model H2}) and a broken power-law (\textbf{model H3}) \cite{cardillo14}.
For leptons, we used a simple power-law with a high energy cut-off in all hadronic models. We fix only the parameters for which we have solid observational evidence: the average medium density, $n=300\, cm^{-3}$, and the radio spectral index, $p'=1.74$; all the other parameters,  such as the normalization constants, $K_{p}$ and $K_{e}$, and the cut-off and break energies, $E_{c}$ and $E_{br}$, are free.
Our best hadronic model is the model H3 with the following proton distribution:
\begin{equation}
\frac{dN_{p,3}}{dE}=\left\{ \begin{array}{ll}
K_{p,1}  \left( \frac{E}{E^p_br} \right)^{-p_{1}} & \textrm{if $E<E_{br}$}\\
K_{p,2}  \left( \frac{E}{E^p_br} \right)^{-p_{2}} & \textrm{if $E>E_{br}$}.
\end{array} \right.
 \label{broken_2}
\end{equation}
This is characterized by an index $p_{1}=2.2\pm0.1$ (for $E<E_{br}$), $p_{2}=3.2\pm0.1$ (for $E>E_{br}$), and an energy break $E^{p}_{br}=20$~GeV. The leptonic contribution to this model is given by a simple power-law for the electrons with $p'=1.74$, and $E^{e}_{c}=12$~GeV (see Fig.~\ref{hadronic}). This model provides a proton energy $W^{p}=5\times10^{49}$ erg and requires an average magnetic field in the emission region, $B=210$ $\mu$G.\\ 


\section{Discussion}
\label{Discussion}
  \begin{figure*}[!ht]
  \centering
   \subfigure{\includegraphics[bb=0 0 805 337,scale=0.35]{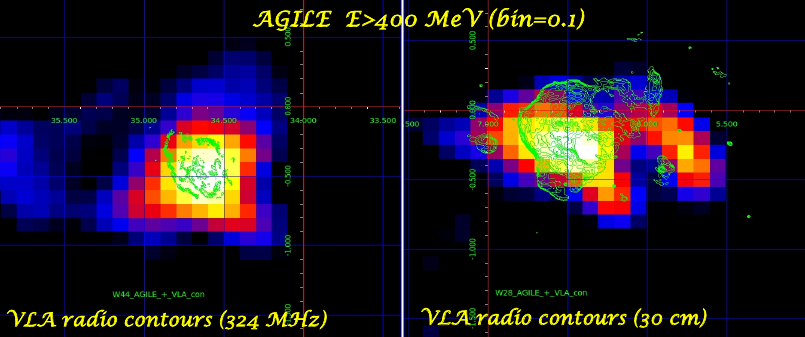}}\\
   \subfigure{\includegraphics[bb=0 0 792 346,scale=0.36]{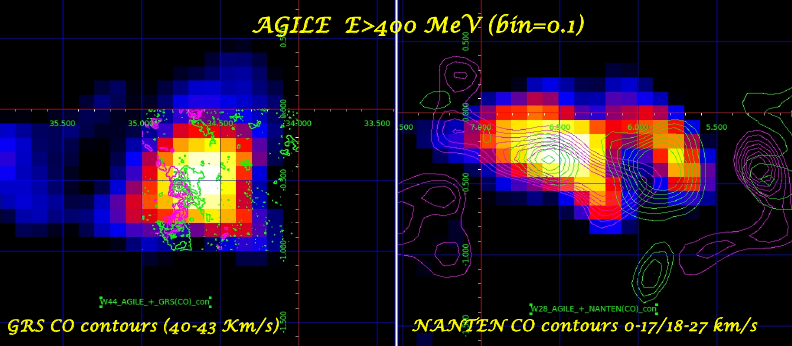}}
   \caption{AGILE gamma-ray maps of the SNR~W44 (left) and the SNR~W28 (right) for E>400 MeV and with binsize=$0.1^{\circ}$, in galactic coordinates.
   In the top panel, radio contours from VLA are overlapped; in the bottom, CO contours from GRS for W44 and from NANTEN fro W28 \cite{cardillo12}.}
   \label{fig_W44W28}
  \end{figure*}
\subsection{W44 main characteristics}
\label{W44 main characteristics}
The most important physical characteristics of the SNR W44 can be summarized in the following way \cite{cardillo14}:
 \begin{itemize}
  \item \textbf{Neutral pion signature:} W44 is the first SNR clearly showing the so-called
  ``pion bump'' that we expect at $E\geq 67$ MeV from $\pi^{0}$-decay photons. The low-energy spectral index, $p_{1}\sim 2.2$ could be affected by the fact that we estimate it in the energy space. In order to confirm (or not) the low-energy behavior we need to consider the SNR~W44 system in the momentum space;
  \item \textbf{High density of the surrounding environment:} We determined that the average density in the W44 shell is  $n_{av} \sim 300 \, \rm cm^{-3}$ with $n\geq 10^{3}$~cm$^{-3}$, which corresponds with CO peaks (see medium panels in  Fig.~\ref{maps}). This feature was also found in other middle-aged  SNRs, like W51c and IC443 \cite{koo10,castelletti11_IC443} and explains the high gamma-ray flux detected from these sources. 
  \item \textbf{High magnetic field:} In W44 our best hadronic models imply a magnetic field  $B\geq 100$ $\mu$G, which is lower than the post-shock magnetic field estimated from Zeeman splitting in the OH masers by \cite{claussen97}, and substantially higher than the equipartition magnetic field \cite{castelletti07}. In most of SNRs, magnetic field estimations give values $B\sim10-10^{2}$ $\mu$G that are
  much higher than the average diffuse galactic value [e.g., see \cite{morlino12} for Tycho, \cite{koo10} for W51c,
 and \cite{tavani10_IC443} for IC443]. This is hardly surprising since magnetic field compression due to the shock interaction with the
 ISM leads to its amplification. We need to then consider a non-linear scenario with a back-reaction of the accelerated particle at
 the shock \cite{bell01}.The large value for the magnetic field in W44 may be linked to the environment density value, $n_{av}\sim 300 \, \rm cm^{-3}$ given by NANTEN2. For a lower density value, we notice that we can enhance the electron density   and make plausible a lower magnetic field \cite{cardillo14}. 
  \item\textbf{ Steepness of the high energy index:} As in \cite{abdo10_W44}, G11, and A13, W44 shows a spectral index $p_{2}\sim3$ for energies above 1~GeV, that is steeper than the values found in other middle-aged SNRs. Alfv\`en damping in a dense environment \cite{malkov11_W44} is a
 mechanism for explaining this behavior, but other possibilities exist  \cite[e.g.,][]{blasi12a,blasi12b}. This is a point requiring deeper investigations in the future.
\end{itemize}

\subsection{W44 and W28: comparison}
\label{W44 and W28: comparison}
\begin{table*}[!ht]
\scriptsize 
\caption{W44 and W28 parameters}
\centering
\label{table_W44W28}
\begin{tabular}{|l|c|c|c|c|c|c|c|c|}
\hline
\textbf{SNR (l,b)}        &\textbf{distance}     & \textbf{age}        &\textbf{radio}    & \textbf{GeV}       &   \textbf{TeV}         & \textbf{n}              & \textbf{B}      &\textbf{MC}\\
                          &\textbf{kpc}        & \textbf{yrs}        &   \textbf{index}          &    \textbf{index}   &     \textbf{index}     & \textbf{$cm^{-3}$}      & \textbf{$\mu G$}&  \\             
\hline
                          &                       &                     &                        &                     &                         &                         &                 &\\
\textbf{W44}(G34.7-0.4)   &  3.1                  & $\sim$20,000        &    0.37                &     $2.4 - >3.0$     &           -             &      250-300            &    $\geq$100    &   yes \\   
                          &                      &                     &                        &                     &                         &                         &                 &\\
\hline
                          &                      &                     &                        &                     &                         &                         &                 &\\
\textbf{W28}(G6.71-0.05)  &1.8-3.3               &  35,000-45,000      &    0.35                &    $\sim 2.6-2.7$    &      $\sim 2.5$        &    5                    &   $10^{2}-10^{3}$&  yes \\
                          &                       &                     &                        &                     &                         &                         &                 &\\
\hline	
\end{tabular}
\end{table*}
At the light of the SNR~W44 characteristics and of the SNR/CR context, we want to present a comparison between this very important source and another important SNR detected in the gamma-ray band, both at GeV and TeV energies, SNR W28 \cite{giuliani10_W28,abdo10_W28,aharonian08_W28}. Even W28 is a middle-aged mixed morphology SNR with dimensions of the shell radio very similar to the W44 ones.
In spite of this, from Table~\ref{table_W44W28} and from Fig.~\ref{fig_W44W28}, we can see that these two remnants have some very different features that lead to a different interpretation of their gamma-ray spectrum.\\
First of all, from the upper panel of Fig.~\ref{fig_W44W28}, we can see that W44 gamma-ray emission has a very good correlation with its radio shell, differently by W28 where no correspondence there is between gamma-ray emission and radio shell.
Observing gamma-ray/CO emission correlation (bottom panel of Fig.~\ref{fig_W44W28}), in W44 the MC seems to be embedded in the remnant; in W28, instead, two different MCs are perfectly correlated with GeV and TeV peaks.
This is confirmed also by estimated average densities of the two SNR shells (see Table~\ref{table_W44W28}); very small for W28 and of order of $10^{2}$ for W44. Moreover, no TeV emission was detected from W44, differently from W28.
This could be explained by the fact that in the case of W44, gamma-ray emission comes from both MC and the SNR shell; TeV particles could be
are already escaped by the remnant. W28 is older than W44 and this implies that the most part of GeV and TeV CR particles are escaped from the remnant.
Consequently, no emission is detected correlated with the shell but only with the two MCs \cite{cardillo12}.\\
This difference is fundamental for their spectral behavior interpretation. We have seen that for W44 a simple linear DSA model fails because high
energy spectral-index is $\sim 3$ when linear DSA model provides an index of $2.6-2.7$. For this reason, we need to consider all possible non-linear mechanisms that could explain the steepening of the spectrum.
In the case of W28, instead, a simple linear DSA model can easily explain its spectral behavior. In \cite{giuliani10_W28}, we find
that where is the peak of GeV emission there is a minimum in the TeV emission, and viceversa. This behavior reflects also in the spectrum; we
can divided it into two components. One from the East cloud, where there is the GeV maximum, and the other from the West cloud complex, where is the TeV maximum \cite{aharonian08_W28,giuliani10_W28}. From \cite{giuliani10_W28}, considering the two clouds at different distances, we can explain gamma-ray data from W28 with the simple energy dependent diffusion, with $\delta\approx0.5$ and $D_{0}=10^{26}$ cm$^{2}$s$^{-1}$ (Bohm diffusion regime).\\
In light of these considerations, it is clear that even in W28 the effect of some non-linear mechanisms affect the system;
consequently, considering only the diffusion mechanism in order to understand its characteristic seems a oversimplification.
However, the low average density of the SNR shell, and the absence of MCs embedded within it exclude most of the non-linear mechanisms
considered so far (see Section~\ref{Theoretical challenges}).

\subsection{W44 and the other gamma-ray emitting SNRs}
\label{Theoretical challenges}
\begin{figure*}[!ht]
 \centering
 \subfigure{\includegraphics[scale=0.45,bb=0 0 497 344]{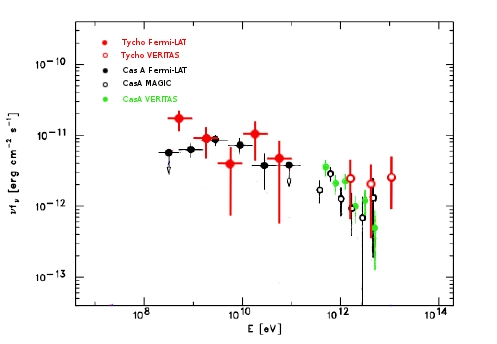}}
 \subfigure{\includegraphics[bb=0 -10 192 134,scale=1.0]{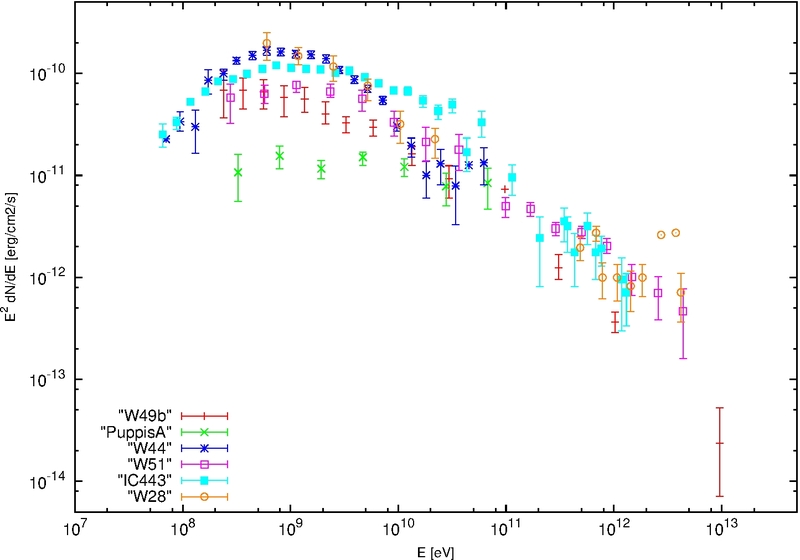}}
 \caption{\textbf{Left Panel}:GeV and TeV spectral points of the two young SNRs Cas~A and Tycho \cite{abdo10_CasA,acciari10_CasA,giordano12_tycho,acciari11_tycho}. \textbf{Right Panel}: GeV and TeV spectral points of the most important middle-aged SNRs: W49b \cite{abdo10_W49b}, Puppis~A \cite{hewitt12_PuppisA},
 W44 \cite{ackermann13_W44,cardillo14}, W51c \cite{aleksic12_W51,abdo09_W51}, IC443 \cite{ackermann13_W44,acciari09_IC443} and W28 \cite{abdo10_W28, aharonian08_W28}}
 \label{fig_SNRspectra}
\end{figure*}
Now we extend the previous comparison at the most important young and middle-aged SNRs emitting in the gamma-ray band. 
The importance of young SNRs is related to two fundamental issues. First, their spectra are not affected by propagation effect as in the case
of middle-aged SNRs. Consequently, analysis of their spectral behavior allows to understand injection spectral index and physical processes that
could have an impact on acceleration mechanism. On the other hand, their young age enhances the chance to detect emission from particles with energies up to $E\sim10^{15}$ eV, called Pevatrons, one of the direct signature for CR acceleration.\\
Gamma-ray data collected by satellites and instruments, however, put us in front of a very challenging reality.
Fig.~\ref{fig_SNRspectra} shows gamma-ray emission from Cas~A and Tycho at GeV and TeV energies.\\
The first problem is that young SNR fluxes are quite faint in the gamma-ray energy band. This leads to the detection of a very low
number of these objects; consequently, it is very difficult have some reliable conclusions about their general behavior.
From theoretical considerations, young SNRs should to have a low energy flux at GeV energies, but a high energy flux in the TeV energy range.
Even if the faintness of the detected gamma-ray flux is strongly related to the SNR distance, there is another parameter that can affect it intrinsically; this is the density value in the SNR surroundings. 
Both Cas~A and Tycho SNRs expand into a low average density medium \cite{hwang12_casA,gomez12_Tycho}, as well as the other young SNRs, even if in all cases are detected some MCs. If, on one side, this low average density can explain a low gamma-ray flux, on the other side, it is not so easy to explain the gamma-ray emission from these SNRs with a (most favorable) hadronic model. An explanation for this behavior is given in \cite{caprioli11b}; where the progenitor wind residual is considered as the CR target. Another explanation could be linked to proton trapping; in a Bohm-like diffusion regime the proton escape time is much greater than the time needed to cross the system lengths.\\ 
However, the most important challenge derived from experimental data is their spectral index, $p=2.3$-$2.4$ \cite{abdo10_CasA,giordano12_tycho}.
Emission at the ``knee'' energies is expected by young SNRs because their emission is not affected by propagation effects. However, no sources were observed at  $E>10$ TeV. Consequently, steepening of young SNR spectrum implies no detection of Pevatrons; in the case of Cas~A and RX J1713-3946, the presence of  high-energy cut-off enhances the system complexity.\\
Differently by young SNRs, middle-aged SNRs have spectra that are influenced by propagation effects. When we analyze their gamma-ray emission,
we have to consider the modification of injection spectral index due to the diffusion. Clearly, it is more difficult to detect the first phases
of the acceleration process and we expect to be difficult to detect Pevatrons because high energy particles are already escaped from these remnants. Detection at TeV energies is possible only in the presence of a target not embedded in the remnant, like in the SNR~W28.\\
In Fig.~\ref{fig_SNRspectra} (right panel), we collected all available GeV and TeV data of most of the middle-aged SNRs studied so far.
The majority of these SNRs have a GeV flux quite high and are easily detectable. The lower flux of W49b and W51c is probably due to their great distances from us with respect to the other remnants). In all cases we detect MC complexes interacting with remnants and magnetic fields have large values.\\ 
Puppis~A seems to be a unique case. It has a low magnetic field and no MC are detected in their surroundings \cite{hewitt12_PuppisA,dubner13_puppisA}.
Its GeV flux is low, even if its distance is not so large (2 kpc), probably due to the absence of a dense target. Moreover, its spectrum has an index $\alpha\sim 2.1$, harder than all the other middle-aged SNRs and also than young SNRs.\\
The middle-aged SNRs gamma-ray spectral indices are in a range $2.6\leq\alpha\leq3$, and, in the oldest SNRs, radio spectral index is harder than the one expected from modifications due to shock waves \cite{castelletti07, onic13}. This implies an electron index, not only harder than $\alpha\sim2$, but also different from the proton one. All these SNRs seem to have similar surroundings and similar characteristics, such as high magnetic field, presence of MCs and so on. In spite of these facts, their gamma-ray spectral indices show that different physical mechanisms are at work. The comparison between W44 and W28 was an example (Section~\ref{W44 and W28: comparison}).\\ 
In order to explain SNR spectral behavior, different from the theoretical expectations, several physical processes are considered so far:
\begin{itemize}
 \item \textbf{Neutrals ``return flux''}: there is a suppression of the Mach number and compression ratio due to the formation of a shock precursor \cite{draine93,blasi12c_neutrals,ohira12};
 \item \textbf{Scattering Center Velocity}: there is the formation of a CR-induced precursor. Low-energy particles feel the lower compression factor at the subshock  and the spectrum becomes steeper. This effect disappears at $E>$ few GeV because high energy particles feel the whole precursor \cite{blasi12c_neutrals};
 \item \textbf{Alfv\'en Damping}: due to the presence of neutrals in the SNR surroundings, it leads to a suppression of scattering center velocity in a certain  energy range and, consequently, of the acceleration efficiency \cite{kulsrud69,malkov11_W44}.
\end{itemize}
All these processes have solid physical reasons but most of them depend on poorly known parameters that, often, are considered separately.
In a complex system like a SNR, we should consider all possible physical processes together, and their mutual interaction. Only in this way
we can have a correct picture of the system.


\section{Conclusions}
\label{Conclusions}

The SNR W44 is a crucial source providing important information about the CR origin in our Galaxy.
However, several characteristics of this SNR, which have been deduced by a multifrequency approach (gamma-ray spectral indices, large
magnetic field), are challenging. 
W44 is a relatively close and quite  bright gamma-ray source. Therefore, an excellent characterization of its gamma-ray spectrum in the range 50-200 MeV has been possible because of the good statistics achieved by AGILE and Fermi-LAT. A re-analysis of the AGILE data from new and updated archives, revisiting radio and CO data of W44, shows the unlikeliness of leptonic-only models in their most natural form: electron distributions constrained by radio data, cannot fit the broad-band W44 spectrum inside a 1-zone model. 
On the other hand, we find that both gamma-ray and radio data can be successfully modeled by different kinds of hadronic models (H1, H2, and H3).\\
The best one is a broken power-law with $p_{1}=2.2 \pm 0.1$ for $E<E_{br}$ and $p_{2}=3.2 \pm 0.1$ for $E>E_{br}$.
Our results regarding the spectral properties of the accelerated proton/ion population by the W44 shock qualitatively agree with the results of \cite{GiuCaTa11}. Source morphology resulted to be different from the previous one. Consequently, the interpretation of its surroundings and of the origin of the gamma-ray emission is different. However, we confirmed the interaction between the remnant and a MC,
probably embedded in the source; most of the gamma-ray emission originates from this region. This can be explained by the age of the SNR. W44 is a middle-aged SNR and most of the high-energy particle diffused far away from the source and this is also the reason of the non detection of TeV emission. In the MC, high density reduces the diffusion and lower-energy particles are trapped inside it, emitting GeV $\gamma$-ray from $\pi^{0}$ decay that we can detect yet.
Independently from the hadronic model used, one big issue generated by Fermi-LAT and AGILE modeling is a photon spectral index, $p=3.0\pm0.1$, steeper than the one provided by theoretical models and the steepest between all the other SNR indices. Moreover, we find a high value of the magnetic field (of order of $10^{2}$ $\mu$G) that appears to be strictly correlated with the surrounding medium density. We see this feature in every SNR known so far, in spite of different surroundings or spectral behavior. 
Placing the case of SNR~W44 in the general SNR/CR context, with a direct comparison with the other SNRs, in a direct way with SNR~W28, we stress the necessity of a deeper knowledge of each SNR and its surroundings.














\end{document}